\documentclass[twocolumn,prb,showpacs,preprintnumbers,amsmath,amssymb]{revtex4}

\usepackage{graphicx}

\begin{document}

\title{Experimental magnetic form factors in Co$_3$V$_2$O$_8$: A combined study of {\it ab initio} calculations, magnetic Compton scattering and polarized neutron diffraction}

\author{N. Qureshi}

\email[Corresponding author:~]{navidq@st.tu-darmstadt.de}

\affiliation{Institute for Materials Science, University of
Technology, Petersenstrasse 23, D-64287 Darmstadt, Germany}

\affiliation{Institut Max von Laue-Paul Langevin, 6 rue Jules
Horowitz, BP 156, 38042 Grenoble Cedex 9, France}

\author{M. Zbiri, J. Rodr\'iguez-Carvajal, A. Stunault, E. Ressouche, T.C. Hansen, M.T. Fern\'andez-D\'iaz and M.R. Johnson}

\affiliation{Institut Max von Laue-Paul Langevin, 6 rue Jules
Horowitz, BP 156, 38042 Grenoble Cedex 9, France}

\author{H. Fuess}

\affiliation{Institute for Materials Science, University of
Technology, Petersenstrasse 23, D-64287 Darmstadt, Germany}

\author{H. Ehrenberg}

\affiliation{Institute for Complex Materials, IFW Dresden, D-01069
Dresden, Germany}

\author{Y. Sakurai and M. Itou}

\affiliation{Japan Synchrotron Radiation Research Institute (JASRI),
SPring-8, Sayo, Hyogo 679-5198, Japan}

\author{B. Gillon}

\affiliation{Laboratoire L\'eon Brillouin (CEA-CNRS), CE Saclay,
91191 Gif-sur-Yvette, France}

\author{Th. Wolf}

\affiliation{Research Center Karlsruhe, Institute of Solid State
Physics, D-76021 Karlsruhe, Germany}

\author{J.A. Rodr\'iguez-Velamazan and J. S\'anchez-Montero}

\affiliation{Instituto de Ciencia de Materiales de Arag\'on (CSIC -
Universidad de Zaragoza), 50009 Zaragoza, Spain}

\affiliation{Institut Max von Laue-Paul Langevin, 6 rue Jules
Horowitz, BP 156, 38042 Grenoble Cedex 9, France}

\begin{abstract}
We present a combination of {\it ab initio} calculations, magnetic
Compton scattering and polarized neutron experiments, which
elucidate the density distribution of unpaired electrons in the
kagome staircase system Co$_3$V$_2$O$_8$. {\it Ab initio} wave
functions were used to calculate the spin densities in real and
momentum space, which show good agreement with the respective
experiments. It has been found that the spin polarized orbitals are
equally distributed between the $t_{2g}$ and the $e_g$ levels for
the spine (s) Co ions, while the $e_{g}$ orbitals of the cross-tie
(c) Co ions only represent 30\% of the atomic spin density.
Furthermore, the results reveal that the magnetic moments of the
cross-tie Co ions, which are significantly smaller than those of the
spine Co ions in the zero-field ferromagnetic structure, do not
saturate by applying an external magnetic field of 2 T along the
easy axis {\it a}, but that the increasing bulk magnetization
originates from induced magnetic moments on the O and V sites. The
refined individual magnetic moments are
$\mu$(Co$_c$)=1.54(4)~$\mu_B$,
$\mu$(Co$_s$)=2.87(3)~$\mu_B$,
$\mu$(V)=0.41(4)~$\mu_B$,
$\mu$(O1)=0.05(5)~$\mu_B$,
$\mu$(O2)=0.35(5)~$\mu_B$ and
$\mu$(O3)=0.36(5)~$\mu_B$ combining to the same
macroscopic magnetization value, which was previously only
attributed to the Co ions.

\end{abstract}

\pacs{75.25.+z; 75.30.Et; 75.30.Gw}

\maketitle

\section{Introduction}
\label{sec:Introduction}

\begin{figure}
\includegraphics[width=1.5in]{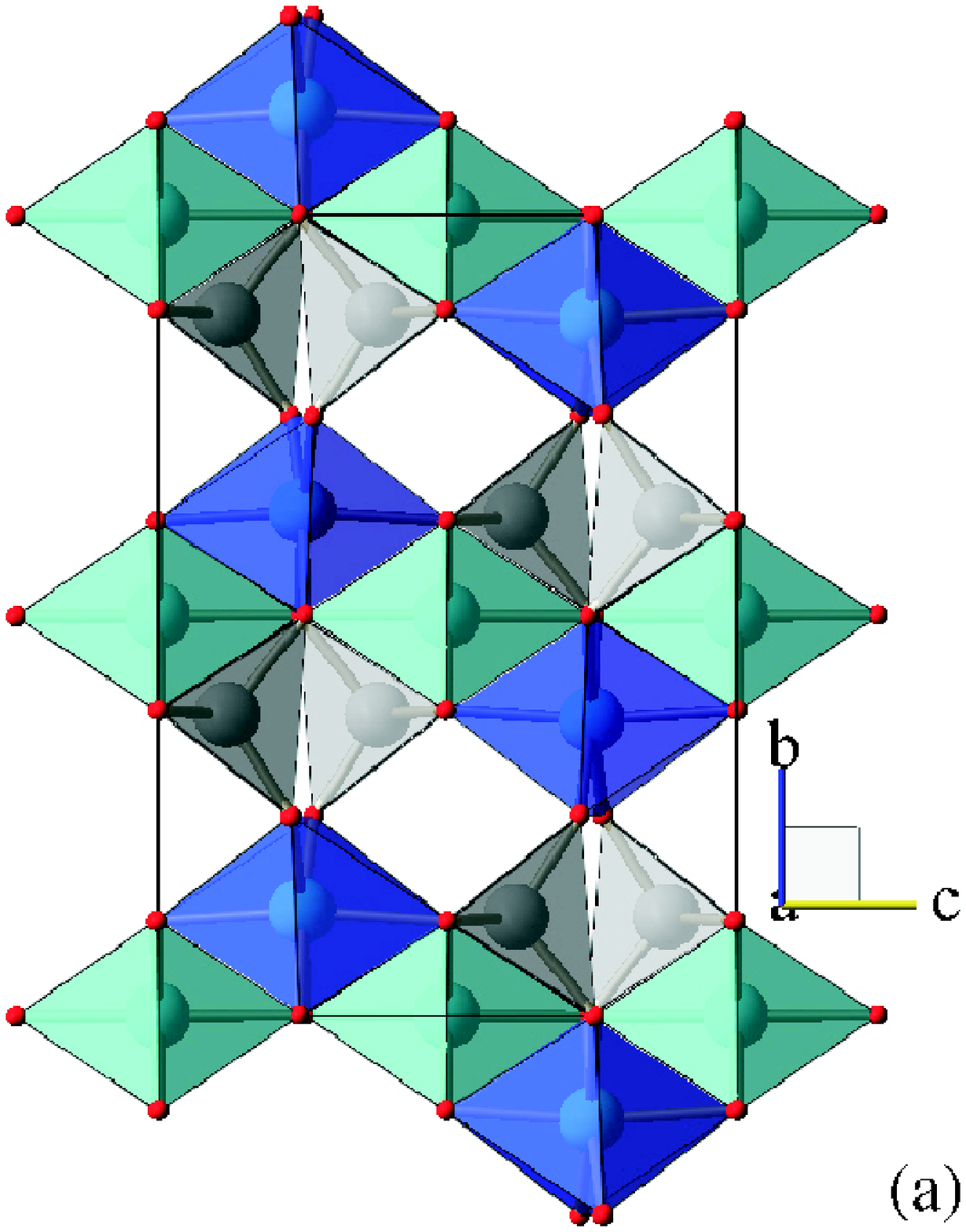}
\includegraphics[width=1.8in]{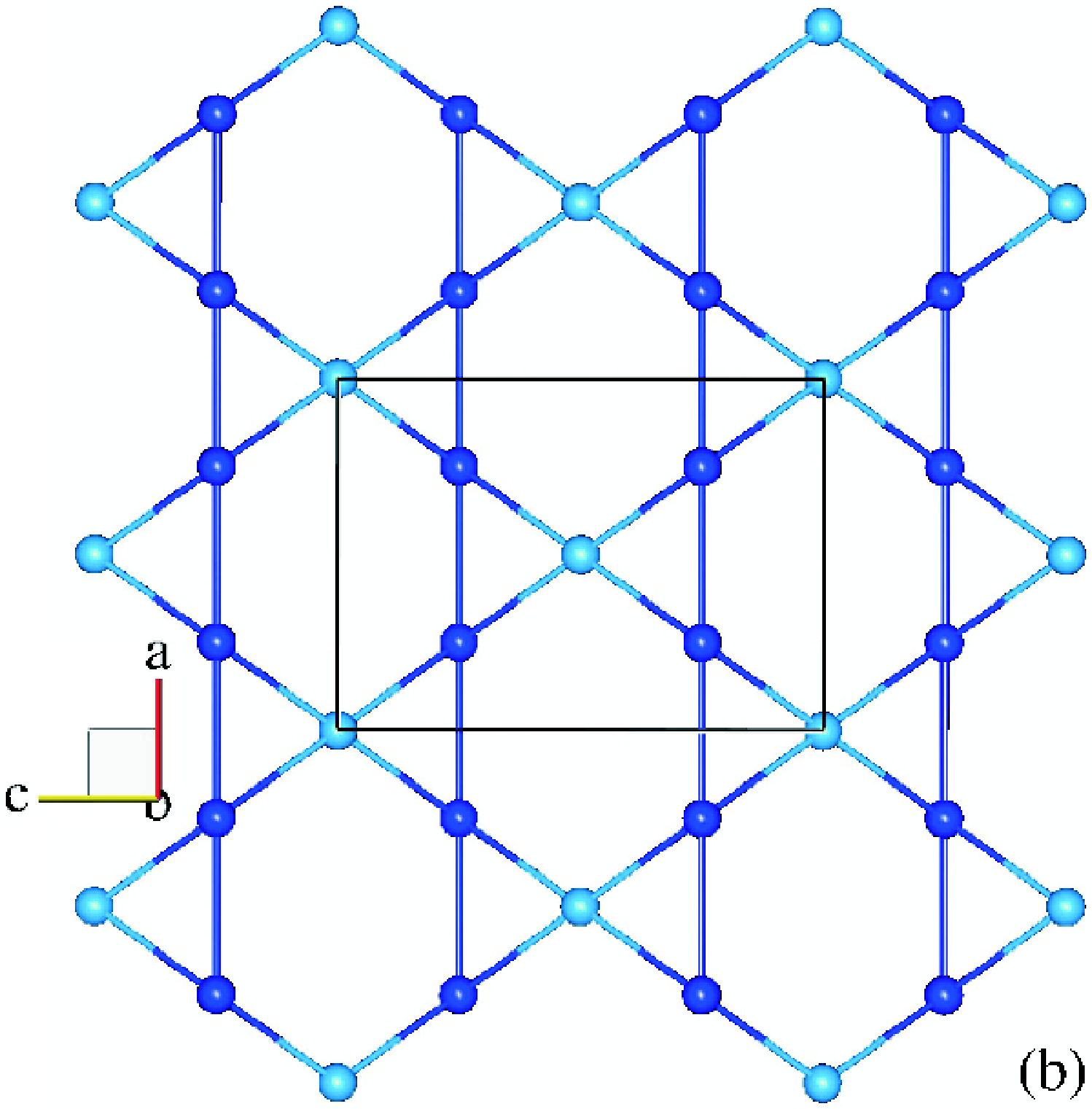}
\caption{\label{fig:structure} (Color online) Visualization of the
kagome staircase structure of Co$_3$V$_2$O$_8$. (a) shows the edge
sharing MO$_6$ octahedra [Co$_c$O$_6$ light blue (light gray),
Co$_s$O$_6$ dark blue (dark gray)] isolated by non-magnetic VO$_4$
tetrahedra (gray). (b) depicts a single kagome staircase viewed
along the {\it b} axis with only the magnetic ions on both
crystallographic sites [Co$_c$ light blue (light gray), Co$_s$ dark
blue (dark gray)].}
\end{figure}

Co$_3$V$_2$O$_8$ represents the 3d transition metal
ortho-oxo-vanadates labeled as kagome staircase structures and
crystallizes in the orthorhombic space group
Cmca.\cite{fue1970,sau1973} Its crystallographic structure is
characterized by edge-sharing CoO$_6$ octahedra forming buckled
layers of corner-sharing triangles, the kagome staircases, which are
separated along the {\it b} axis by VO$_4$ tetrahedra
(Fig.~\ref{fig:structure}).
The magnetic exchange is effectuated by a 90$^\circ$ Co-O-Co
intralayer pathway. Interestingly, this system exhibits a sequence
of five magnetic phase transitions featuring temperature dependent
incommensurate antiferromagnetic phases with commensurate lock-ins
and a ferromagnetic ground state, where all magnetic structures are
collinear along the {\it a} axis.\cite{che2006} The ferromagnetic
structure reveals two strongly different magnetic moments for the
cross-tie Co ions (4a site) and spine Co ions (8e site) of 1.54
$\mu_B$ and 2.73
$\mu_B$,\cite{che2006} respectively, despite the fact
that both Co$^{2+}$ ions apparently present high-spin configurations
as macroscopic measurements exhibit saturation of the cross-tie
moments.\cite{wil2007/2} In the kagome staircase structure several
nearest and next-nearest neighbor exchange interaction pathways are
possible,\cite{che2006} which is the motivation for this study.
Investigating the magnetization density, which is reflected by the
respective magnetic form factors of the ions, may reveal preferred
exchange pathways with the presence of magnetization on the involved
O sites. Eventual induced magnetization on the empty d-shell of V
sites would allow interlayer coupling by super-superexchange.
Therefore, magnetic Compton scattering and polarized neutron
diffraction experiments have been carried out leading to the spin
density in momentum space and the magnetization density in real
space, respectively. However, detailed and precise analysis of these
quantities are required in order to determine the exact contribution
of the atomic species involved in the studied system. Quantum
chemical modeling is therefore needed to gain insights, at a
molecular level, into the electronic structure of the two
cobalt-oxide octahedra. In this context, {\it ab initio} cluster
calculations were performed for the Co$_c$O$_6$ and Co$_s$O$_6$
octahedra yielding precise molecular orbitals (MO) and wave
functions (wf). The latter were used to analyze the experimentally
observed density distributions. The refinement of the contribution
of each MO/wf, at a quantum chemical level, to the real and momentum
space densities simultaneously is a powerful procedure allowing
interesting and valuable features of the magnetic form factors to be
determined.

\section{Ab initio calculations}
\label{sec:abinitio}

The two different clusters Co$_c$O$_6$ and Co$_s$O$_6$ were modelled
separately. The calculations were performed within the framework of
the Kohn-Sham formulation of the density functional theory using the
PC GAMESS program.\cite{pcgamess} The functional B3LYP was employed
to approximate the exchange-correlation interaction. The B3LYP is a
hybrid functional, well adapted to study transition metal compounds
and magnetic interactions, in which a predefined amount of the exact
Hartree-Fock exchange is added to the well known pure density
functionals.\cite{b3lyp1,b3lyp2,b3lyp3,b3lyp4} The atoms in the
cluster were described using Ahlrich's pVDZ atomic orbital (AO)
basis set\cite{sch1992}
Co(14s,8p,5d,1p)/[5s,2p,2d,1p],O(7s,4p,1d)/[3s,2p,1d]. The notations
(klm) and [klm] indicate the number of Gaussian type orbitals and
contracted Gaussian type orbitals, respectively. In order to mimic
the Madelung potential, the two quantum mechanical clusters were
surrounded by point charges (PC) according to the Effective Fragment
Potential method.\cite{gor2001} As previously reported for other
systems,\cite{rad2005,pas2006,bar1988,pas1993} the choice of the
embedding method was shown to be crucial for the physical meaning of
the {\it ab initio} calculations. To avoid the electron density
leaking out of the cluster, a boundary region has been introduced,
which is formed by effective core potentials (ECP) placed in the
nearest cationic positions around the cluster. Thus, the first
coordination shell of Co$^{2+}$ and V$^{5+}$ ions has been described
by ECPs according to the SBKJC ECP basis set.\cite{ste1992} As
native ECPs for Co and V do not treat the 3s and 3p electrons as
core electrons, those ECPs would in fact be too compact due to the
higher number of valence electrons. In order to overcome this
problem the ECPs for Mg and Al have been used due to the fact that
their ionic radii are closer to those of Co and V, respectively.
Except for the 3d shells, the remaining electrons are replaced by an
effective potential. Thus, 1565 PC (3x3x3 unit cells with the
respective Co ion in the center) have been built to mimic the
Madelung potential on the cluster. The MO coefficients relevant to
the Co3d were extracted from the simulations and used to model the
Magnetic Compton Profiles (MCP) and the magnetic form factors.

\section{Experimental}
\label{sec:Experimental}

\subsection{Unpolarized neutron diffraction}

Co$_{3}$V$_2$O$_8$ single crystals have been grown from self-flux in
a ZrO$_2$/Y crucible by the slow cooling method. Some of the
crystals have been ground and investigated at the high-flux neutron
powder diffractometer D20 (Institut Laue Langevin, Grenoble)
confirming the absence of parasitic phases. The nuclear structure of
a chosen single crystal has been studied at the four-circle
diffractometer D9 (Institut Laue Langevin, Grenoble). A set of more
than 500 independent reflections up to $\sin\theta/\lambda$=0.92 has
been measured using two different wavelengths in order to determine
the extinction effects ($\lambda_1$=0.835~\AA{},
$\lambda_2$=0.512~\AA{}). The data collection has been performed in
the paramagnetic phase at $T$=13.5 K, which is just above the N\'eel
temperature of 11.2 K. In order to reveal possible structural
changes between the paramagnetic and the ferromagnetic phase due to
structural phase transitions or magnetostriction an additional
nuclear structure investigation has been carried out at the single
crystal diffractometer D15 (Institut Laue Langevin, Grenoble) under
the same experimental conditions as the polarized neutron
diffraction experiment (Sec.~\ref{sec:pnd}), i.e. at $T$=3.5 K with
an applied magnetic field of $H$=2 T along the easy axis {\it a} and
using a wavelength of $\lambda$=0.854~\AA{}.

\subsection{Polarized neutron diffraction (PND)}
\label{sec:pnd}

The real space magnetization density has been studied at the hot
neutron spin polarized two-axes diffractometer 5C1 (Laboratoire
L\'eon Brillouin, Saclay). Neutrons from the source are
monochromated and polarized by the (111) reflection of a magnetized
Heusler crystal Cu$_2$MnAl. The wavelength is 0.84~\AA{}, which
corresponds to the maximum flux of the hot source and is ideal for
studying large domains of reciprocal space. The polarization factor
of the beam is $p=-0.88$. In order to fully magnetize the sample and
avoid beam depolarization a magnetic field of $H$=2 T has been
applied along the easy axis {\it a}. The flipping ratios $R$
(Eq.~\ref{eq:rfl}), the ratios between the spin-up and spin-down
intensities, of over 500 independent $(hkl)$ reflections with
$h$=0,1,2 have been measured in the ferromagnetic phase at $T$=3.5
K.
\begin{equation}
R=\frac{(F_N^2+q^2F_M^2)p^+_p+2F_Nq^2F_Mp^+_m+(1-q^2)q^2
F_M^2y_{pm}}{(F_N^2+q^2F_M^2)p^-_p+2F_Nq^2
F_Mp^-_m+(1-q^2)q^2F_M^2y_{pm}}, \label{eq:rfl}
\end{equation}
$F_M$ and $F_N$ denote the magnetic and nuclear structure factors.
$q$=$\sin\alpha$ is a geometric factor with $\alpha$ being the angle
between the scattering vector and the magnetization vector. The
parameters $p^\pm_{p/m}$ and $y_{pm}$ are extinction correction
factors of the respective cross-sections according to
Ref.~\onlinecite{bon1976}.

\subsection{X-ray magnetic Compton scattering (MCS)}

The investigation of the spin density in momentum space has been
carried out at the High Energy Inelastic Scattering beamline BL08W
at SPring-8 in Hyogo, Japan. This beamline is designed for magnetic
Compton scattering spectroscopy as it offers high energy
elliptically polarized X-rays emitted from an Elliptic Multipole
Wiggler. The incident photon beam with an energy range of 170-300
keV is monochromated and focused by an asymmetric Johann type
monochromator using the Si (620) reflection. The sample
magnetization is achieved with a superconducting magnet with a
maximum field of 2.5 T and a minimum polarity-switching time of 5
seconds. The backscattered photon energy is analyzed by a
10-segmented Ge solid state detector positioned at a scattering
angle of 178.4$^\circ$. The experiment has been carried out with an
incident photon energy of 176.3 keV, which gives a good compromise
between the beam intensity and the scattering cross section. The
initial interest of applying this method to Co$_3$V$_2$O$_8$ was to
map the momentum space spin density of the ferromagnetic phase as a
projection onto the {\it b$^*$-c$^*$} plane in order to gather
information about the 3d electron spin states and to correlate the
results with those obtained from the polarized neutron diffraction
experiment, but the experimental conditions and especially the large
magnetic anisotropy of the system did not allow that. The minimal
achievable sample temperature is approximately 5.6 K, i.e. close
below the magnetic transition into the antiferromagnetic phase. It
can be seen in the magnetic phase diagrams of
Co$_3$V$_2$O$_8$\cite{qur2007,wil2007/2} that at this temperature
already weak magnetic fields applied along the {\it b} or {\it c}
axis induce a magnetic phase transition into the antiferromagnetic
phase, while $H||{\it a}$ stabilizes the ferromagnetic one. The
necessity of applying a magnetic field of considerable strength and
therewith magnetizing the sample along the incident beam in order to
increase the magnetic contribution to the scattering cross-section
beside the requirement of turning the sample about a vertical axis
to be able to reconstruct the two-dimensional momentum density, led
to a change of strategy. To make sure not to induce magnetic phase
transitions by rotating the sample with respect to the field
direction the measurements have been carried out within the
antiferromagnetic phase at $T$=7.5 K and by applying a magnetic
field of $H$=2 T with the induced ferromagnetic component lying in
the {\it b-c} plane. In addition to the trivial directions [010] and
[001] four further directional MCPs $J_{mag}$ in the $p_y$-$p_z$
plane have been investigated. An additional profile has been
measured along the [100] direction by decreasing the applied
magnetic field to 0.25 T. A directional MCP yields the projection of
the spin momentum space density onto the scattering vector, which is
by definition $p_z$:
\begin{equation}
J_{mag}(p_z)=\int\limits_{-\infty}^{\infty}\int\limits_{-\infty}^{\infty}|\chi_\uparrow(\mathbf
p)|^2-|\chi_\downarrow(\mathbf p)|^2dp_xdp_y
\end{equation}
with $\chi_{\uparrow(\downarrow)}(\mathbf p)$ denoting the momentum
wf of an occupied majority (minority) spin state. The Compton
profiles of the respective sample magnetization states have been
recorded for 60 seconds, repeating the cycle $[+--+-++-]$ multiple
times.

\section{Results}

\subsection{Nuclear structure and extinction}

The nuclear structure investigation at 13.5 K on the four-circle
diffractometer D9 confirmed the correct phase formation of the
orthorhombic structure (space group Cmca) with $a$=6.015(3)~\AA{},
$b$=11.480(5)~\AA{} and $c$=8.289(4)~\AA{}. The observed integrated
intensities have been corrected for absorption by applying the
transmission factor integral $\exp[\mu(\bar{t}_{in} +
\bar{t}_{out})]$ and analyzed by simultaneously fitting a structure
model to both datasets with $\lambda_1$ and $\lambda_2$ using
FullProf.\cite{fullprof} The refinement process included the atomic
positions and isotropic temperature factors of Co and O plus three
additional extinction parameters according to an anisotropic
Shelx-like empirical extinction correction.\cite{lar1970} The atomic
position and temperature factor of V have been fixed in all
refinements, because of its low coherent neutron scattering cross
section. The refined structural parameters ($R$=5.3) are listed in
Tab.\ref{tab:nuc}.\newline
\begin{table}[htbp]
\caption{\label{tab:nuc}Structural parameters of the investigated
Co$_3$V$_2$O$_8$ single crystal.}
\begin{ruledtabular}
\begin{tabular}{ccccc}
Atom&{\it x}&{\it y}&{\it z}&{\it B}(\AA{}$^2$)\\
\hline
Co1 & 0 & 0 & 0 & 0.27(10)\\
Co2
& 0.25 & 0.1328(7) & 0.25 & 0.20(7) \\ V & 0 & 0.3773 & 0.1204 & 0.30 \\
O1 & 0 &
0.2489(3) & 0.2700(4) & 0.38(4)   \\ O2 & 0 & 0.0008(4) & 0.2448(4) & 0.33(4) \\
O3 & 0.2702(3) & 0.1185(3) & 0.9990(2) & 0.33(4)\\ \\
\multicolumn{5}{c}{Extinction parameters}\\
\multicolumn{5}{c}{$x_{11}$=1.0(1)  $x_{22}$=0.36(5)  $x_{33}$=0.6(1) }\\
\end{tabular}
\end{ruledtabular}
\end{table}
In order to analyze the nuclear structure under identical conditions
as the flipping ratio measurement ($T$=3.5 K, $H$=2 T), only those
reflections have been measured for which the ratio $\gamma$ between
the magnetic and the nuclear structure factor has been derived from
the observed flipping ratios. The ferromagnetic contribution to the
integrated intensity $I$ has been canceled out according to
\begin{equation}
I\sim F_N^2+|\mathbf Q_M|^2=F_N^2+q^2F_M^2=F_N^2(1+q^2\gamma^2),
\end{equation}
where $\mathbf Q_M$ is the magnetic interaction vector. The
subsequent refinement showed that no considerable change in the
nuclear structure has taken place, i.e. the derivation of the $F_M$
from observed flipping ratios by using the observed $F_N$ at
$T=13.5$ K is justified.\newline

\begin{figure}
\includegraphics[width=3in]{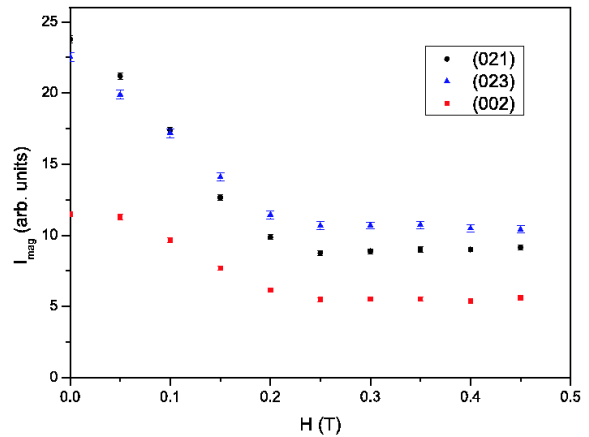}
\caption{\label{fig:magext} (Color online) Intensity of three
different magnetic reflections in dependence of an applied magnetic
field revealing primary extinction effects.}
\end{figure}

All low-angle nuclear reflections suffer considerably from
extinction, therefore, special attention has been paid to the
extinction of magnetic scattering. As the flipping ratio treatment
uses the same extinction parameters for both nuclear and magnetic
scattering, it is important to verify, if the extinction effects are
indeed comparable. Therefore, three strong magnetic reflections have
been measured as a function of applied magnetic field after the
sample has been cooled in zero-field to 3.5 K. Fig.~\ref{fig:magext}
shows the integrated intensities of three reflections after the
nuclear contribution has been subtracted. The field dependence of
the magnetic contribution reveals a surprising and interesting
tendency: Instead of increasing with increasing applied field, as
one would expect if the cross-tie moments get saturated, the
intensity of magnetic scattering drops significantly.
This observation can be explained with field dependent increase of
primary extinction. At $H$=0 T the sample exhibits a multidomain
state with presumably negligible extinction effects. By increasing
the field the magnetic domains grow until they reach approximately
the size of the structural domains. On reaching saturation at
$H$$\approx$0.25 T the primary extinction effects for magnetic
scattering should be comparable to those of nuclear scattering. The
mosaicity which governs secondary extinction should a priori not be
affected. In order to verify these assumptions the extinction
correction factor $y$ has been calculated for three magnetic
reflections using the refined extinction parameters from the nuclear
structure refinement according to the anisotropic FullProf model:
\begin{equation}
y=\left[1+\frac{0.001F_M^2\lambda^3(x_{11}h^2+x_{22}k^2+x_{33}l^2)}{4\sin(2\theta)(\sin\theta/\lambda)^2}\right]^{-\frac{1}{2}}
\end{equation}
The calculated values have been compared with the observed ones,
which can easily be deduced from the intensity ratios at $H$=0 T and
$H$=0.25 T. The results are listed in Tab.~\ref{tab:magext}. It can
be seen that the calculated extinction factors are to a greater or
lesser extent comparable with the observed ones. Nevertheless, the
extinction of magnetic scattering seems to be underestimated.
\begin{table}[htbp]
\caption{\label{tab:magext}Observed and calculated extinction
correction parameters for three low-angle magnetic reflections.}
\begin{ruledtabular}
\begin{tabular}{ccccc}
(hkl) & $\sin\theta/\lambda$ (\AA$^{-1}$)& $F_{M,obs}$ (10$^{-12}$ cm)  & $y_{obs}$ & $y_{cal}$\\
\hline
(021) & 0.10595 & 4.87(1) & 0.39(7) & 0.46(3) \\
(002) & 0.12064 & 3.39(1) &  0.47(6) & 0.64(3) \\
(023) & 0.20083 & 4.72(1) &  0.47(6) & 0.60(3) \\
\end{tabular}
\end{ruledtabular}
\end{table}

\subsection{Real space magnetization density}

\begin{figure}
\includegraphics[width=3in]{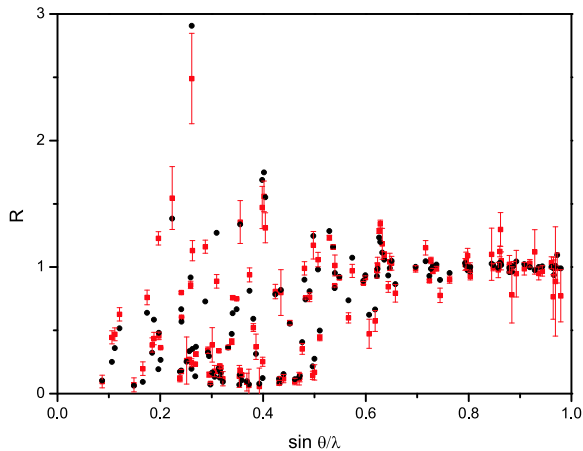}
\caption{\label{fig:rvssintl} (Color online) Observed (squares) and
calculated (dots) flipping ratios as a function of $\sin
\theta/\lambda$.}
\end{figure}

\begin{figure}
\hspace{0.35cm}\includegraphics[width=3.02in]{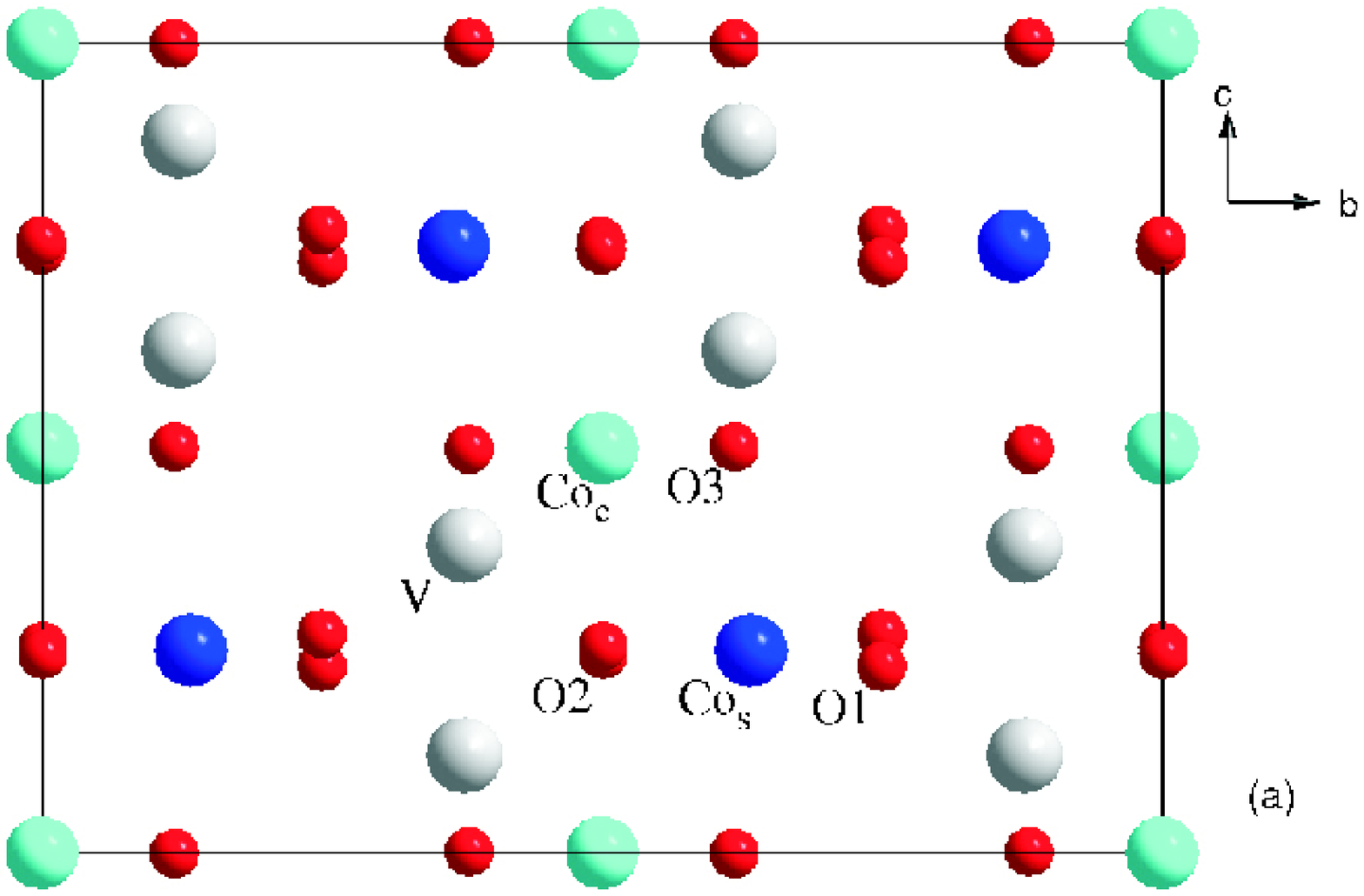}
\includegraphics[width=3.4in]{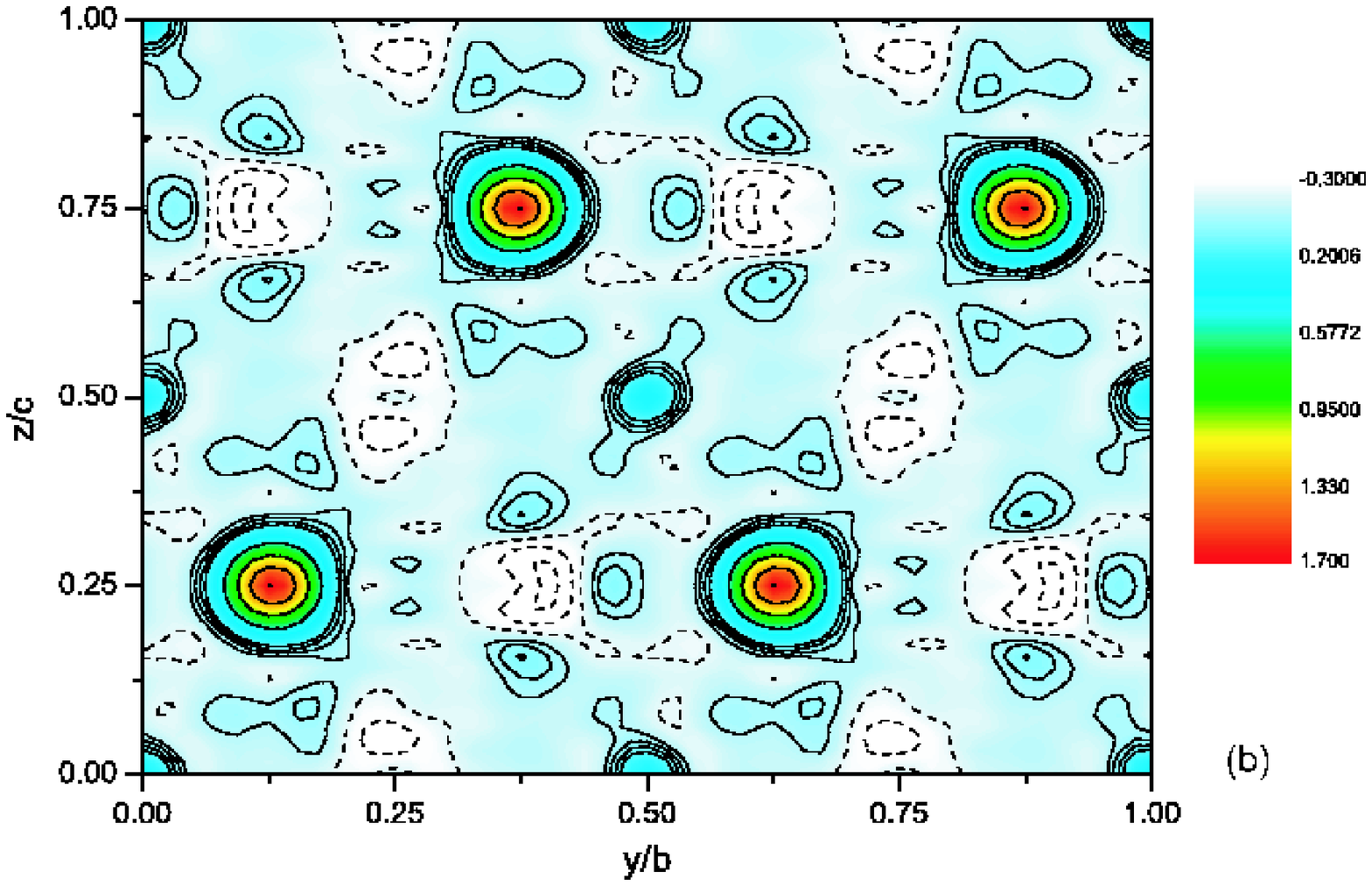}
\includegraphics[width=3.4in]{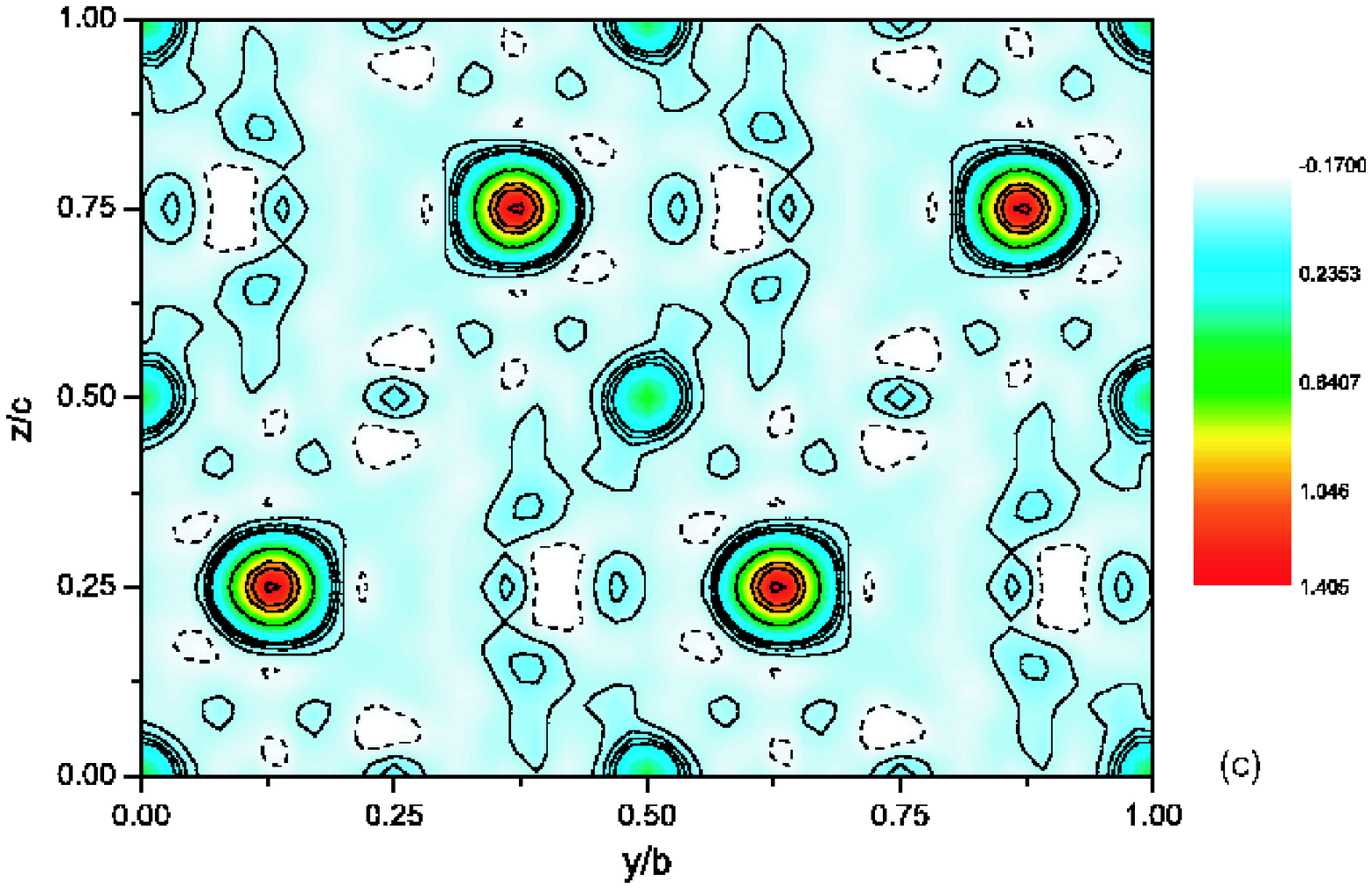}
\caption{\label{fig:real} (Color online) (a) Crystal structure
viewed along the {\it a} axis. (b) Experimental and (c) calculated
magnetization density as a projection onto the {\it b-c} plane.
Contour lines defining positive values are drawn as solid lines in
0.05 $\mu_B$/\AA{}$^2$ intervals between 0
$\mu_B$\/\AA{}$^2$ and 0.15
$\mu_B$/\AA{}$^2$ and in 0.4
$\mu_B$/\AA{}$^2$ intervals above. Negative
isodensities are represented by broken lines in 0.1
$\mu_B$/\AA{}$^2$ steps.}
\end{figure}

The nuclear structure factors, which have been deduced from the
unpolarized neutron experiments, have been used to derive the
magnetic structure factors from the observed flipping ratios by
solving Eq.~\ref{eq:rfl} with respect to $F_M$. The individual
observed and calculated flipping ratios are shown in
Fig.~\ref{fig:rvssintl} as a function of $\sin \theta/\lambda$.
As the crystal structure is centrosymmetric the experimental
magnetization density can directly be reconstructed by a Fourier
synthesis. Fig.~\ref{fig:real}(b) shows the projection of the
observed magnetization density onto the {\it b-c} plane, while
Fig.~\ref{fig:real}(a) depicts the unit cell viewed along the {\it
a} axis in order to assign the density peaks to the respective
atoms. As it has been assumed in the previous section, the Co$_c$
moments do not get saturated, but significant magnetization density
is present on V and O sites. While the density is quite localized
for the V, O1 and O2 sites, rather diffuse density can be observed
around the O3 site. The split density peaks of O1 result from the
fact, that actually two O1 ions are visible in the projection.
Similarly, the density around the Co$_s$ ions seems to be much
higher compared to the Co$_c$ ions, which is due to the fact, that
two Co$_s$ ions are contained in the projection, while only one
Co$_c$ ion is projected. Besides the superexchange pathways
Co$_s$-O2-Co$_c$ and Co$_s$-O3-Co$_c$ an interlayer exchange becomes
evident with the non-zero magnetization density of V and O1. Further
{\it ab initio} solid state computations are currently being
performed to simulate the spin density map of Co$_3$V$_2$O$_8$ and
to elucidate the composite mechanisms of the induced magnetic
moments on the different O and V sites. This will be the subject of
a forthcoming publication.\cite{zbi2008}

\subsection{Momentum space density}

\begin{figure}
\includegraphics[width=2.8in]{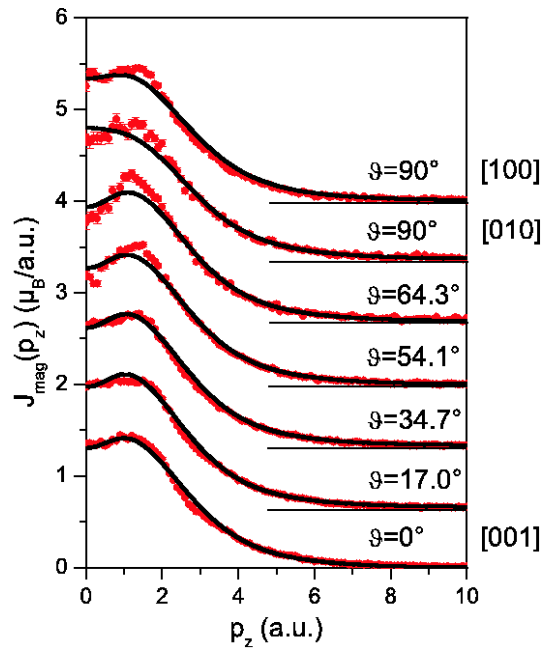}
\caption{\label{fig:mcp} (Color online) Observed (dots) and
calculated (solid lines) normalized directional MCPs (shifted
vertically in order to improve clarity, horizontal lines serve as a
guide for the eye). The abscissa $p_z$ is taken to be parallel to
the respective scattering vector. $\vartheta$ denotes the angle
between a respective MCP and the [001] direction.}
\end{figure}

\begin{figure}
\includegraphics[width=3.2in]{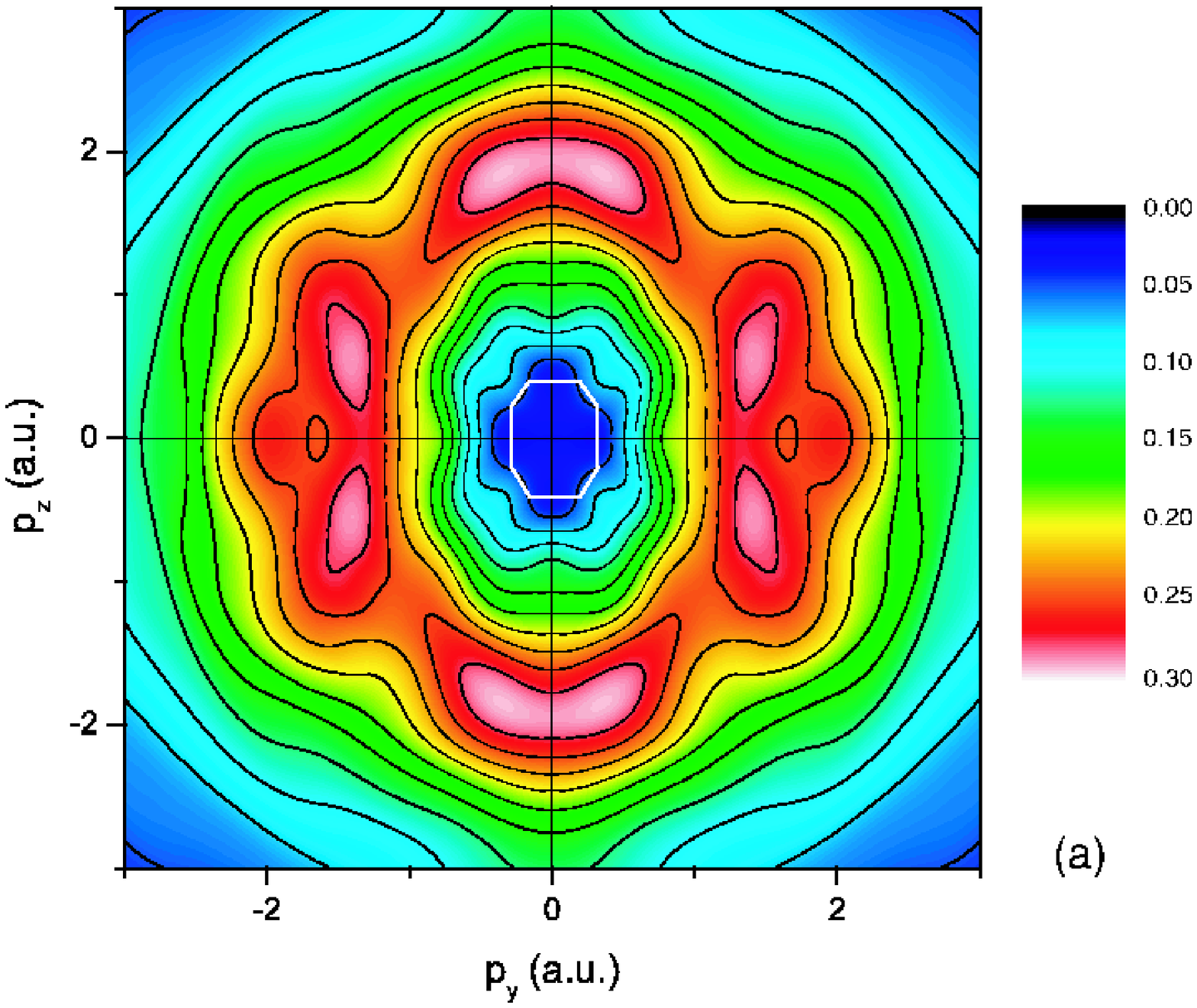}
\includegraphics[width=3.2in]{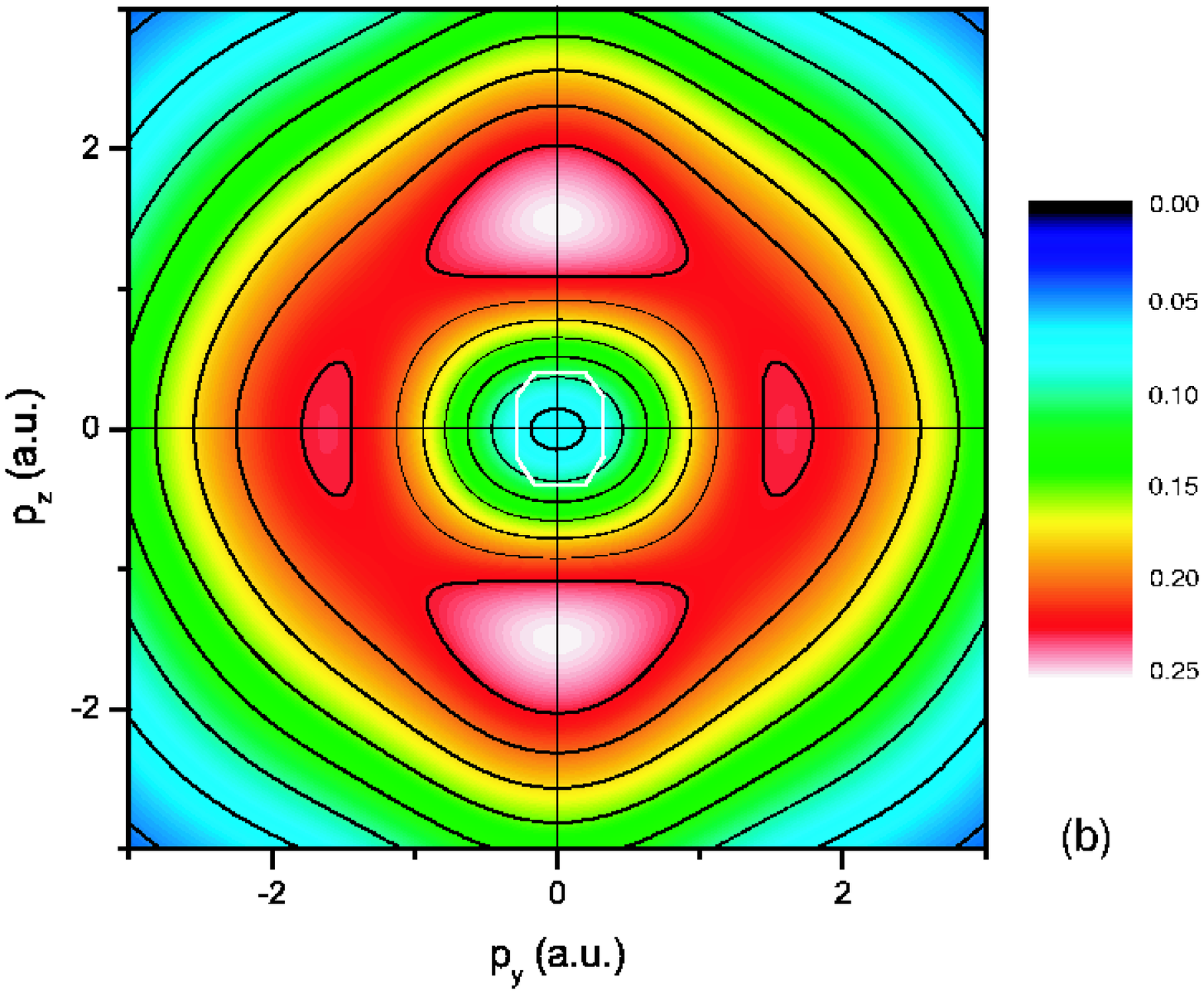}
\caption{\label{fig:mom} (Color online) Reconstructed experimental
(a) and calculated (b) spin momentum density in the $p_y$-$p_z$
plane. Contours are drawn in 0.025 $\mu_B$/(a.u.)$^3$
intervals. White solid lines depict the boundary of the first
Brillouin zone.}
\end{figure}

The MCPs were extracted by taking the difference of the scattered
intensities $I^+$ and $I^-$ of the respective charge Compton
profiles. Due to the fact that the more intense charge Compton
profiles still exhibit relatively large values at the outermost
measured positions of $p_z=\pm 10$ a.u. (atomic units), the actual
number of electrons between -10 a.u. and +10 a.u. is evaluated from
the profiles interpolated using tabulated data for the elements
resulting from Hartree-Fock calculations.\cite{big1975} Before
summing up the magnetic intensity of each detector cell, the data
have been corrected for the detector cell efficiency, sample
absorption and scattering cross-section according to
Ref.~\onlinecite{zuko}. Furthermore, the energy scale of each
detector cell has been calibrated by measuring a radioactive sample
with well known emission energies. \\
The experimental MCPs were folded at $p_z$=0 to increase statistical accuracy by taking the average of
each branch. The area under each profile has been normalized to the
number of magnetic electrons per formula unit. With the use of iron
standards the induced ferromagnetic component can be deduced from
the magnetic effect, which is the relative contribution of the MCP
to the total Compton profile
\begin{equation}
M_0=\frac{I^+-I^-}{I^++I^-}\cdot 100\%.
\end{equation}
The magnetic effects of the respective directional MCPs are listed
in Tab.~\ref{tab:mageff} with their corresponding ferromagnetic
components induced parallel to the scattering vector. In the
$p_y$-$p_z$ plane the spin moments monotonically decrease with
increasing angle between the magnetic field direction and the
$[001]$ direction, which directly reveals the magnetic anisotropy
with {\it b} being the hard axis of this
system.\cite{bal2004,wil2007/2} Effective beam path length dependent
multiple scattering effects due to the rotation of the sample are
estimated to be less than a few percent of the spin values given in
Tab.~\ref{tab:mageff}.
\begin{table}[htbp]
\caption{\label{tab:mageff}Magnetic effects of the respective
directional MCPs with a magnetic field $H$ applied along the
scattering vector. $\vartheta$ denotes the angle between a MCP and
the $[001]$ direction.}
\begin{ruledtabular}
\begin{tabular}{ccccc}
MCP& $\vartheta$ ($^\circ$)&$H (T)$&$M_0$ (\%)& S ($\mu_B$)\\
\hline
$[001]$ & 0 & 2 & 0.541 & 0.616 \\
$[023]$ & 17 & 2 & 0.488 & 0.556 \\
$[012]$ & 34.7 & 2 & 0.415 & 0.472 \\
$[011]$ & 54.1 & 2 & 0.274 & 0.312 \\
$[032]$ & 64.3 & 2 & 0.229 & 0.261 \\
$[010]$ & 90 & 2 & 0.095 & 0.108 \\
$[100]$ & 90 & 0.25 & 0.251 & 0.287 \\
\end{tabular}
\end{ruledtabular}
\end{table}
Fig.~\ref{fig:mcp} shows the normalized observed MCPs, which reveal
similar shapes for the seven investigated crystallographic
directions.
Using all profiles except the one along the [100] direction the
two-dimensional momentum spin density in the $p_y$-$p_z$ plane has
been reconstructed by the direct Fourier-transform
method.\cite{suz1989,tan2001} The calculation has been performed on
a grid with a distance of 0.1 a.u. between each point. The result is
shown as a two-dimensional contour plot in Fig.~\ref{fig:mom}(a).
Low spin density can be recognized inside the first Brillouin zone
(BZ), which extends beyond its border along the
$\langle$010$\rangle$ and $\langle$001$\rangle$ directions. In the
vicinity of the first BZ border the density increases more rapidly
with increasing momentum along $\langle$021$\rangle$. Peaks are
present at $(p_y,p_z)$=$(0.35,1.85)$ and $(1.4,0.55)$.

\subsection{Correlated refinement in both spaces}

\begin{figure}
\includegraphics[width=3in]{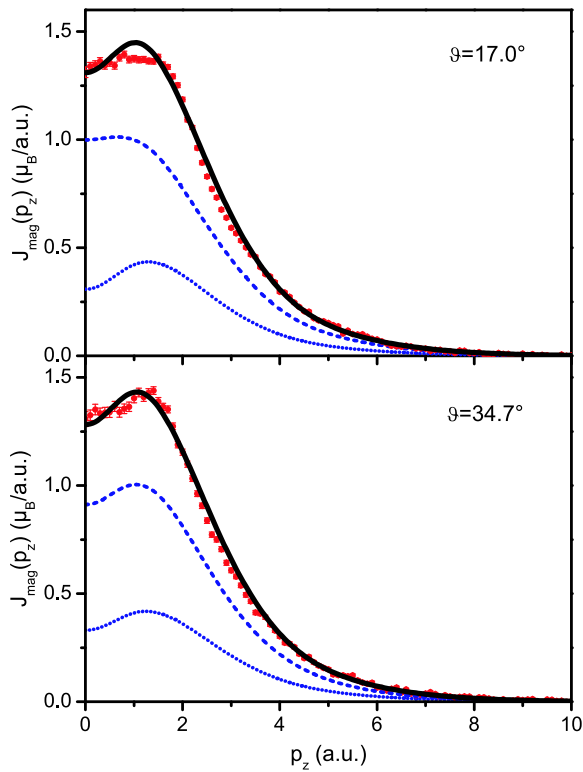}
\caption{\label{fig:contr} (Color online) Observed (dots) and
calculated (solid lines) MCPs along two different directions. The
dotted and dashed lines show the contribution of the Co$_c$O$_6$ and
the Co$_s$O$_6$ cluster, respectively.}
\end{figure}

The idea behind correlating the density distributions in real and
momentum space is that the population of each spin polarized orbital
must be represented likewise in the observed densities in both
spaces. The fact that the MCS method only samples the spin part and
the PND method both the spin and the orbital part of the magnetic
moment has been handled in the following way. As the observed MCPs
have been normalized, the area under each profile corresponding to
the size of the spin moment is not refined. The refined parameters
were the populations of each spin polarized orbital (in correlation
with the real space quantities), thus only the shape of each profile
is refined. The refined magnetic moments stem solely from the PND
data. In order to analyze the observed MCPs theoretical ones have
been calculated by projecting the square of the {\it ab initio} wf
onto the respective scattering vector. Thereby the symmetry
relations between the different cluster density distributions in the
unit cell have to be taken into account, which yields two and four
symmetrically inequivalent Co$_c$O$_6$ and Co$_s$O$_6$ clusters,
respectively (Tab.~\ref{tab:clsym}).
\begin{table}[htbp]
\caption{\label{tab:clsym}Symmetry relations between the two and
four inequivalent Co$_c$O$_6$ and Co$_s$O$_6$, respectively.}
\begin{ruledtabular}
\begin{tabular}{cccc}
cluster& Co position & symmetry relation to c$_1$/s$_1$\\
\hline
$c_1$& $(0,0,0)$ & $xyz$ \\
$c_2$ & $(0,\frac{1}{2},\tfrac{1}{2})$ & $xy\bar{z}$ \\
$s_1$ & $(\frac{1}{4},y,\frac{1}{4})$ & $xyz$ \\
$s_2$ & $(\frac{3}{4},y,\frac{1}{4})$ & $\bar{x}yz$ \\
$s_3$ & $(\frac{3}{4},\bar{y},\frac{3}{4})$ & $\bar{x}\bar{y}\bar{z}$ \\
$s_4$ & $(\frac{1}{4},\bar{y},\frac{3}{4})$ & $x\bar{y}\bar{z}$ \\
\end{tabular}
\end{ruledtabular}
\end{table}
The point symmetries of the Co$_c$O$_6$ and Co$_s$O$_6$ clusters are
$2/m..$ and $.2.$,\cite{tables} which correspond to $2/m..$ and
$.2/m.$ in momentum space. Due to the special symmetry of the
Co$_s$O$_6$ density, the projections of the different clusters in
momentum space are invariant for the principal axes and those in the
$p_y$-$p_z$ plane. In the case of the Co$_c$O$_6$ clusters, the
projections onto non-principal axes in the $p_y$-$p_z$ yield
different profiles, which need to be averaged. The projected
orbitals have been convoluted with a Gaussian function having a full
width at half maximum of the instrumental resolution. As reported
previously\cite{koi2001} the fact that the projection of each MO in
momentum space has a characteristic shape makes it possible to
refine its population $\beta_k$ which is the contribution to the
observed MCP:
\begin{equation}
J_{mag}(p_z)=\int\limits_{-\infty}^{\infty}\int\limits_{\infty}^{\infty}\sum_{k}\beta_k\chi^2_k(\mathbf
p)dp_xdp_y
\end{equation}
Here, $\chi_k(\mathbf p)$ denotes a momentum space MO/wf. Similarly,
the populations $\beta_k$ of the real space MOs can be used to
deduce the magnetic form factors $f_X(\mathbf q)$ of the respective
elements $X$ by calculating the Fourier transform of the atomic spin
density:\
\begin{equation}
f_X(\mathbf
q)=\int\limits_{-\infty}^{\infty}\int\limits_{-\infty}^{\infty}\int\limits_{-\infty}^{\infty}\sum_{k}\beta_k\psi^2_{k,X}(\mathbf
r)\exp(2\pi i\mathbf q \mathbf r)d\mathbf r,
\end{equation}
where $\psi_{k,X}$ defines the real space MO $k$ only including the
atomic orbitals $\phi_{i,X}$ of element $X$=Co, O. With this
procedure the observed flipping ratios can be refined based on a
simple aspheric magnetic form factor model deduced from {\it ab
initio} wf. For the V ions the analytic approximation of the
V$^{4+}$ form factor\cite{lis1971} has been used. Refining the
population parameters for each MCP individually yields excellent
agreement with the observed profiles. But since the refinement
process exhibits numerous local minima with significantly varying
results, it has been considered more reasonable to include all MCPs
in the refinement despite the magnetic anisotropy. The respective
population parameters $\beta_k$ have been refined simultaneously in
both spaces together with the magnetic moments of Co, V and O by
minimizing the function
\begin{align}
\chi^2=&\frac{1}{2}\sum_{i}\frac{(R_{i,obs}-R_{i,cal})^2}{\sigma^2_{i,obs}}\notag \\
+&\frac{1}{2}\sum_{n}\sum_{j}\frac{[J_{n,obs}(p_{z,j})-J_{n,cal}(p_{z,j})]^2}{\sigma^2_{j,obs}}
\end{align}
with $i$ and $j$ defining discrete data points of the PND and MCS
experiment, respectively, and $n$ referring to the respective MCPs.
The refinement yields fairly good agreement expressed by
$R_{MCS}$=5.7 and $R_{PND}$=9.6 for the respective experiments. The
refined total magnetic moments along the {\it a} axis are
\begin{align}
\mu(\text{Co}_c)&=1.54(4)~\mu_B\notag \\
\mu(\text{Co}_s)&=2.87(3)~\mu_B\notag \\
\mu(\text{V})&=0.41(4)~\mu_B\notag \\
\mu(\text{O}1)&=0.05(5)~\mu_B\notag \\
\mu(\text{O}2)&=0.35(5)~\mu_B\notag \\
\mu(\text{O}3)&=0.36(5)~\mu_B.\notag
\end{align}
Summing the magnetic moments of all ions in the unit cell weighted
by their site multiplicity and dividing by the number of Co ions
yields an averaged magnetization of 3.45
$\mu_B$/Co$^{2+}$. This value shows excellent
agreement with the macroscopic magnetization for $H=2$ T along the
{\it a} axis reported in Ref.~\onlinecite{wil2007/2}. The resulting
relative orbital populations are listed in Tab.~\ref{tab:fitpars}.
The refined parameters were used to calculate the MCPs, which are
depicted as solid lines in Fig.~\ref{fig:mcp}. Fig.~\ref{fig:contr}
shows that the line shapes of the two respective contributing parts
(Co$_c$O$_6$ and Co$_s$O$_6$ MOs) are different concerning the ratio
between the value at the peak and at $p_z$=0 and that they vary with
the projection angle, which is important for a meaningful fit.
\begin{table}[htbp]
\caption{\label{tab:fitpars}Refined orbital occupation parameters of
the Co$_c$O$_6$ and Co$_s$O$_6$ clusters.}
\begin{ruledtabular}
\begin{tabular}{ccc}
orbital& Co$_c$ & Co$_s$\\
\hline $d_{xy}$& 0.27(2) & 0.12(2) \\
$d_{xz}$ & 0.27(2) & 0.12(2) \\
$d_{yz}$ & 0.16(2) & 0.26(2) \\
$d_{x^2-y^2}$ & 0.17(2) & 0.30(2) \\
$d_{3z^2-r^2}$ & 0.13(2)& 0.20(2) \\
\end{tabular}
\end{ruledtabular}
\end{table}
From the calculated MCPs the momentum space spin density has been
reconstructed [Fig.~\ref{fig:mom}(b)]. The calculated real space
magnetization density map [Fig.~\ref{fig:real}(c)] has been obtained
by a Fourier synthesis of the calculated magnetic structure factors.
The main features of the respective density maps coincide well,
although some differences are evident: the dip in the momentum space
density around $p_z$=0 is not pronounced well in the calculated map
and has a shape, which is rotated by 90$^\circ$ with respect to the
observed map. This possibly results from strong hybridization
effects between the Co3d and O2p orbitals. Furthermore, the
disagreements between the experimental and calculated MCPs are most
probably due to the limitation of the cluster calculations.
Considering a larger sized cluster or the solid state, the gap
between the experiment and fitting could be reduced. The real space
spin density of the O1 and O3 sites is slightly underestimated.
Furthermore, density peaks exist next to the spine site density
along the {\it z} axis, which do not coincide with atomic positions.
However, this fact can be attributed to truncation effects in the
Fourier series.

\subsection{Discussion}

The study presented here combining several methods reveals very
interesting magnetic properties of the kagome staircase system
Co$_3$V$_2$O$_8$. The previous assumption, that the ferromagnetic
structure at zero magnetic field is not fully ordered because of the
Co$_c$ only exhibiting 1.54 $\mu_B$, is disproved.
Previous macroscopic magnetization measurements\cite{wil2007/2}
indeed showed a saturated moment of approximately 3.4
$\mu_B$ per Co site at $H$=2 T along {\it a}, but the
results of the polarized neutron single-crystal diffraction
experiment with adequate extinction correction presented here reveal
that the field dependent increase of magnetization stems from the V,
O2 and O3 sites. Hereby, the V and O2 site show quite localized
magnetization density, while the O3 density seems to be smeared out
due to truncation effects in the Fourier series. A periodic {\it ab
initio} calculation confirms the existence of magnetization density
on the V and O sites and will be presented elsewhere.\cite{zbi2008}
The spin polarized density on O2 and O3, which are those oxygen ions
in the Co$_c$O$_6$ clusters, may be a strong indication for a
partially covalent character of the Co$_c$ ions and the reason for
their relatively low magnetic moment compared to Co$_s$. The
magnetization density distribution clearly exhibits the
superexchange pathways between the two different Co sites, but it
indicates also the interlayer coupling, which is mediated by the
V-O1 bridge. Combining the methods of polarized neutron diffraction
and magnetic Compton scattering allowed us to refine the occupations
of the Co3d orbitals in a stable way. Like it has been previously
reported but with inverted values\cite{fue1982} the two
crystallographically different Co ions exhibit different spin
polarized orbital occupations. While the unpaired electrons are
equally distributed between the $t_{2g}$ and $e_g$ levels for the
Co$_s$ ion, the magnetic signal stems by only 30\% from the $e_{g}$
orbitals for the Co$_c$ ion as a consequence of the spin transfer
from the surrounding O ions. Concerning the $e_g$ orbitals of both
ions the basal plane orbital $d_{x^2-y^2}$ is more populated than
the apical $d_{3z^2-r^2}$ orbital. This possibly indicates a higher
exchange interaction between the Co$_s$ ions via an intermediate O2
ion. In the case of the Co$_c$ ions it could be a hint that the
magnetic exchange with the spine Co ions takes place preferentially
via an O3 ion.

\begin{acknowledgments}
This research was supported by the {\it Deutsche
Forschungsgemeinschaft} within the priority program 1178. The
magnetic Compton scattering experiment was performed with the
approval of the Japan Synchrotron Radiation Research Institute
(JASRI) (Proposal No. 2007B2021). Helpful discussions with Dr. A. A.
Granovsky and Prof. A. Koizumi are thankfully acknowledged.
\end{acknowledgments}

\end{document}